\documentclass[12pt]{article}
\usepackage{amsmath,amssymb}

\makeatletter
\@addtoreset{equation}{section}

\makeatother

\usepackage{amsmath,amssymb}

\newcommand{\ga}{\alpha}
\newcommand{\gb}{\beta}
\newcommand{\gc}{\gamma}

\newcommand{\gq}{\theta}

\newcommand{\gk}{\kappa}
\newcommand{\gl}{\lambda}
\newcommand{\gs}{\sigma}

\newcommand{\gw}{\omega}

\newcommand{\gC}{\Gamma}

\newcommand{\gW}{\Omega}

\newcommand{\pr}{\partial}
\newcommand{\mr}{\mathrm}
\newcommand{\mrd}{\mathrm{d}}

\newcommand{\fr}{\frac}
\newcommand{\lb}{\label}

\newcommand{\br}{\overline}

\renewcommand{\(}{\left(}
\renewcommand{\)}{\right)}
\renewcommand{\[}{\left[}
\renewcommand{\]}{\right]}

\begin{document}

\thispagestyle{empty}

{\hbox to\hsize{
\vbox{\noindent July 2007 \hfill hep-th/0702139 v3 }}}

\noindent
\vskip2.3cm
\begin{center}

{\Large\bf CONFORMALLY FLAT FRW METRICS~\footnote{
Supported in part by the Japanese Society for Promotion of Science (JSPS)}}

\vglue.3in

Masao Iihoshi~\footnote{Email address: iihosi-masao@ed.tmu.ac.jp},
Sergei V. Ketov~\footnote{Email address: ketov@phys.metro-u.ac.jp},
and Atsushi Morishita~\footnote{Email address: morishita-atsushi@ed.tmu.ac.jp}
\vglue.2in
{\it Department of Physics\\
     Tokyo Metropolitan University\\
     1--1 Minami-osawa, Hachioji-shi\\
     Tokyo 192--0397, Japan}
\end{center}
\vglue.2in
\vskip1cm
\begin{center}
{\Large\bf Abstract}
\end{center}
\vglue.1in
\noindent We find a new family of non-separable coordinate transformations 
bringing the FRW metrics into the manifestly conformally flat form. Our results
are simple and complete, while our derivation is quite explicit. We also
 calculate all the FRW curvatures, including the Weyl tensor.
  
\newpage

\section{Introduction}

The fundamental Cosmological Principle of the {\it spatially} homogeneous and
isotropic $(1+3)$-dimensional Universe (at large scales) gives rise to the
standard {\it Friedman-Robertson-Walker} (FRW) metrics of the form 
\cite{mtw,ll}
$$  ds^2 =  dt^2 - a^2(t)\left[ \frac{dr^2}{1-kr^2} +r^2d\Omega^2\right]
\eqno(1.1)$$
where the function $a(t)$ is known as the scale factor in  `cosmic' 
coordinates $(t,r,\theta,\phi)$; we use $c=1$ and  
$d\Omega^2= d\theta^2 +\sin^2\theta d\phi^2$, while $k$ is the FRW topology 
index taking values $(-1,0,+1)$. Accordingly, the FRW 
metric (1.1) admits a 6-dimensional isometry group $G$ that is either 
$SO(1,3)$, $E(3)$ or $SO(4)$, acting on the orbits $G/SO(3)$, with  the spatial
3-dimensional sections $H^3$, $E^3$ or $S^3$, respectively.~\footnote{Our
notation follows ref.~\cite{ll}, and it is given in Appendix.}
 
By the coordinate change, $dt=a(t)d\eta$, the FRW metric (1.1) can be rewritten
to the form
$$ ds^2 = a^2(\eta)\left[ d\eta^2 - \frac{dr^2}{1-kr^2} - r^2d\Omega^2\right]
\eqno(1.2)$$ 
which is obviously (4d) conformally flat in the case of $k=0$. It immediately 
implies that the 4d Weyl tensor of the FRW metric vanishes in the
`flat' case of $k=0$. In fact, the FRW Weyl tensor also vanishes for $k=-1$ and
 $k=+1$ (see Appendix for our explicit check). In its turn, it implies that 
there exist the coordinate transformations that bring the FRW metrics to the 
conformally flat form in the non-trivial cases of $k=-1$ and $k=+1$ too.
 
Though the fact that the FRW Weyl tensor vanishes for all topologies is known,
while some special coordinate transformations bringing the FRW metric
 (1.2) to the conformally flat form are also known (see the end of Sec.~2),
to the best of our knowledge, we are not aware of any systematic treatment of 
{\it all} such transformations in the literature, as well as their most  
general form, with an {\it arbitrary~} scale factor. Because of the great 
importance of the FRW metrics to physics, knowing such explicit transformations
 is desirable, when taking advantage of the vanishing Weyl tensor. In this 
paper we find that the transformations in question are surprisingly simple.

Our paper is organized as follows. In Sec.~2 we consider the case of $k=-1$
in great detail. Sec.~3 is devoted to the case of $k=+1$. We find that there
is the fundamental difference between those two cases, as regards the existence
of separable real solutions. Possible physical applications are briefly 
discussed in Sec.~4. In Appendix we summarize our notation, calculate all 
the FRW curvatures, and verify that all the FRW Weyl tensor components vanish.
   
\section{The open FRW case $k=-1$}

Let's introduce a coordinate $\chi$ by
\begin{align}
	r = \sinh \chi 
\end{align}
Then the FRW metric reads
\begin{align}
	-\mrd s^2 = a^2(\eta) \[ -\mrd\eta^2 + \mrd\chi^2 + \sinh^2\chi 
\mrd\gW^2\]
	\lb{2.2}
\end{align}

We are looking for some new local coordinates $\xi(\eta,\chi)$ and 
$R(\eta,\chi)$ 
in which the metric (2.2) would be manifestly conformally flat, i.e.
\begin{align}
	-\mrd s^2 = a^2(\xi,R)A^2(\xi,R)
	\[ -\mrd\xi^2 + \mrd R^2 + R^2\mrd\Omega^2\] \lb{2.3}
\end{align}
where
\begin{align}
	{a}(\xi,R ) := a \( \eta ( \xi,R ) \) \lb{2.4}
\end{align}
and $A$ is yet another function of $\xi$ and $R$. Substituting
\begin{subequations}
	\begin{align}
		\mrd\xi = \xi,_\eta \mrd\eta + \xi,_\chi \mrd\chi \\ \lb{2.5}
		\mrd R = R ,_\eta \mrd\eta + R,_\chi \mrd\chi
	\end{align}
\end{subequations}
into eq.~(\ref{2.3}) gives
\begin{align}
	- \mrd s^2 = {a}^2(\xi,R)A^2(\xi,R) [ 
		&-(\xi,_\eta)^2\mrd\eta^2
		-2\xi,_\eta\xi,_\chi\mrd\eta\mrd\chi
		-(\xi,_\chi)^2\mrd\chi^2 \nonumber \\
		&+(R,_\eta)^2\mrd\eta^2
		+2R,_\eta R,_\chi\mrd\eta\mrd\chi
		+(R,_\chi)^2\mrd\chi^2 \nonumber \\
		&+R^2\mrd\gW^2
	] \lb{2.6}
\end{align}
which should be the same as eq.~(\ref{2.2}), Hence, the functions $A$, $\xi$ 
and $R$ 
obey the following non-linear partial differential equations:
\begin{subequations}
	\begin{align}
		-1 = A^2[-(\xi,_\eta)^2 + (R,_\eta)^2] \lb{2.7a}\\
		1 = A^2[-(\xi,_\chi)^2 + (R,_\chi)^2] \lb{2.7b} \\
		0 = -2 \xi,_\eta \xi,_\chi + 2 R,_\eta R,_\chi \lb{2.7c}\\
		\sinh^2\chi = A^2(\xi,R)R^2 \lb{2.7d}
	\end{align}
\end{subequations}

A substitution of eq.~(\ref{2.7d}) into eqs.~(\ref{2.7a}) and (\ref{2.7b}) 
gives
\begin{subequations}
	\begin{align}
		(\xi,_\eta)^2 &=
		\fr{R^2}{\sinh^2\chi} +(R,_\eta)^2  \lb{2.8a}\\
		(\xi,_\chi)^2 &=
		-\fr{R^2}{\sinh^2\chi} + (R,_\chi)^2 \lb{2.8b}
	\end{align}
\end{subequations}
whereas a substitution of eqs.~(\ref{2.8a}) and (\ref{2.8b}) into 
eq.~(\ref{2.7c}) gives
\begin{align}
	-\fr{R^2}{\sinh^2\chi}
	+(R,_\chi)^2 -(R,_\eta)^2
	=0 \lb{2.9}
\end{align}
Equations (\ref{2.8a}), (\ref{2.8b}) and (\ref{2.9}) now imply 
\begin{subequations}
	\begin{align}
		(R,_\eta)^2 &= (\xi,_\chi)^2 \\
		(R,_\chi)^2 &= (\xi,_\eta)^2
	\end{align}
\end{subequations}
and, hence,
\begin{subequations}
	\begin{align}
		R,_\eta = &\gs_1 \xi,_\chi ~~~~~~~(\gs_1=\pm1)\\
		R,_\chi = &\gs_2 \xi,_\eta, ~~~~~~~(\gs_2=\pm1)
	\end{align}
\end{subequations}

The original equations (\ref{2.7a}) to (\ref{2.7d}) are invariant under the 
sign flip of $\xi$ 
and $R$, so that we can remove one of those sign ambiguities. Let's redefine 
$\xi \rightarrow \gs_1\xi$ in order to get
\begin{subequations}
	\begin{align}
		R,_\eta = & \xi,_\chi \lb{2.12a}\\ 
		R,_\chi = &\gs_3 \xi,_\eta.~~~~~~~(\gs_3:=\gs_1\gs_2=\pm1)
		\lb{2.12b}
	\end{align}
\end{subequations}

When choosing the elliptic case $\gs_3=-1$, eqs.~(\ref{2.12a}) and (\ref{2.12b}) 
are nothing but the Cauchy-Riemann equations, whose general solution for $R$ and 
$\xi$ is given by the real and imaginary parts, respectively, of an arbitrary 
complex (holomorphic) function $F(\eta \pm i\chi)$. However, it is inconsistent 
with the remaining equations (\ref{2.7c}) and (\ref{2.7d}). Therefore, since we 
are interested in a real solution, we have to choose the hyperbolic case, $\gs_3=+1$.

A solution must satisfy the integrability condition
\begin{align}
	R,_\eta,_\chi &= R,_\chi,_\eta \lb{2.13}
\end{align}
which via eqs.~(\ref{2.12a}), (\ref{2.12b}) and (\ref{2.13}) yields a 
{\it linear} (!) equation
\begin{align}
	\(
		\fr{\pr^2}{\pr \eta^2}
		- \fr{\pr^2}{\pr\chi^2}
	\)\xi
	=0 \lb{2.14}
\end{align}

Equation (\ref{2.14}) is just the two-dimensional wave equation 
whose general solution is given by
\begin{align}
	\xi=
	\xi_+(\eta+\chi) + \xi_-(\eta - \chi) \lb{2.1.1}
\end{align}
where $\xi_+$ and $\xi_-$ are arbitrary functions of $\eta+\chi$ and 
$\eta-\chi$, 
respectively. 
Substituting eq.~(\ref{2.1.1}) into eq.~(\ref{2.12a}) yields
\begin{align}
	R,_\eta &=
	\fr{\pr}{\pr\chi}\xi_+ + \fr{\pr}{\pr\chi}\xi_- \nonumber\\
	&=\fr{\pr}{\pr\eta}\xi_+ - \fr{\pr}{\pr\eta}\xi_-
\end{align}
whose integration with respect to $\eta$ gives
\begin{align}
	R = \xi_+ -\xi_- + f(\chi)
\end{align}
where the function $f(\chi)$ is actually a constant because of 
eq.~(\ref{2.12b}). That constant amounts to a trivial shift of $R$, 
while it can also be included into any of the functions $\xi_+$ or $\xi_-$, 
so we simply set it to zero. As a result, we get
\begin{subequations}
	\begin{align}
		\xi &= \xi_+(\eta+\chi) + \xi_-(\eta-\chi) \lb{2.1.4a}\\
		R &= \xi_+(\eta+\chi) - \xi_-(\eta-\chi) \lb{2.1.4b}
	\end{align}
\end{subequations}
with some yet to be determined functions $\xi_+$ and $\xi_-$ of a single 
variable.

To determine the remaining functions, we substitute eqs.~(\ref{2.1.4a}) and 
(\ref{2.1.4b})
 into eq.~(\ref{2.7a}) and get
\begin{align}
	\fr{1}{\sinh^2\chi}=\fr{4\xi'_+\xi'_-}{(\xi_+ -\xi_-)^2} 
	\lb{2.1.5}
\end{align}
where the primes mean the derivatives with respect to $\eta + \chi$ or $
\eta -\chi$, 
respectively. We now introduce the new coordinates $x$ and $y$ as follows:
\begin{subequations}
	\begin{align}
		x :=\eta+\chi&,~~y :=\eta-\chi \lb{2.1.6}\\
		\chi=\fr{x-y}{2}&,~~\eta=\fr{x+y}{2} 
	\end{align}
\end{subequations}
Then (\ref{2.1.5}) takes the form
\begin{align}
	\fr{1}{\sinh^2\fr{x-y}{2}}
	=\fr{4\xi'_+(x) \xi'_-(y)}{(\xi_+(x) -\xi_-(y))^2} \lb{2.1.7}
\end{align}
It should be satisfied for any values of $x$ and $y$ so, when we set $y=0$ in 
(\ref{2.1.7}), we achieve a separation of variables,
\begin{align}
	\fr{1}{\sinh^2\fr{x}{2}} \mrd x
	=\fr{4\xi'_-(0)}{(\xi_+(x)-\xi_-(0))^2}\mrd\xi_+
\end{align}
After integration, we get
\begin{align}
	\xi_+(x)-\xi_-(0)
	=\fr{2\xi'_-(0)}{\coth\fr{x}{2} +c_+} \lb{2.1.9}
\end{align}
where $c_+$ is an integration constant.
Similarly, when we set $x=0$ in eq.~(\ref{2.1.7}), we get
\begin{align}
	\xi_-(y)-\xi_+(0) 
	=\fr{2\xi'_+(0)}{\coth\fr{y}{2} +c_-}
\end{align}
and, when we set $x=y=0$, we get
\begin{align}
	\xi_+(0)=\xi_-(0) \lb{2.1.11}
\end{align}

We can remove some integration constants, without changing a value of $R$ by 
the redefinitions 
$\xi_+(x)\rightarrow\xi_+(x)-\xi_-(0)$ and $\xi_-(y)\rightarrow\xi_-(y) 
-\xi_+(0)$, 
because of 
eq.~(\ref{2.1.11}). Also, when differentiating eq.~(\ref{2.1.9}) with respect 
to $x$ at
 $x\rightarrow 0$, we find
\begin{align}
	\xi'_+(0) = \xi'_-(0)
\end{align}
Hence, after a redefinition $\xi_\pm \rightarrow\fr{1}{2\xi'_+(0)}\xi_\pm$, we 
arrrive  at 
\begin{subequations}
	\begin{align}
		\xi_+(x) &=
		\fr{1}{\coth\fr{x}{2}+c_+} \lb{2.1.13a} \\ 
		\xi_-(y) &=
		\fr{1}{\coth\fr{y}{2}+c_-} \lb{2.1.13b}
	\end{align}
\end{subequations}

Substututing eqs.~(\ref{2.1.13a}) and (\ref{2.1.13b}) into eq.~(\ref{2.1.7})
for a final check, and using an identity  $\sinh(\ga\pm\gb)=
\sinh\ga\cosh\gb\pm\cosh\ga\sinh\gb$, just gives $c_+=c_-:=c$. Collecting all
together, we arrive at our main result
\begin{subequations}
	\begin{align}
		\xi
		&= \fr{1}{\coth\fr{\eta+\chi}{2}+c} + 
\fr{1}{\coth\fr{\eta-\chi}{2}+c} 
		\lb{2.1.14a}\\
		R
		&= \fr{1}{\coth\fr{\eta+\chi}{2}+c} - 
\fr{1}{\coth\fr{\eta-\chi}{2}+c} 
		\lb{2.1.14b}\\
		A^2(\eta,\chi)&= \fr{
			(\coth\fr{\eta+\chi}{2}+c)^2
(\coth\fr{\eta-\chi}{2}+c)^2
		}{
			(\coth\fr{\eta-\chi}{2}-\coth\fr{\eta+\chi}{2})^2
		}\sinh^2\chi \lb{2.1.14c}
	\end{align}
\end{subequations}

The inverse transformation is given by
\begin{subequations}
	\begin{align}
		\eta
		&=\coth^{-1}\[\fr{2}{\xi+R}-c\]
		+ \coth^{-1}\[\fr{2}{\xi-R}-c\] \\
		\chi
		&=\coth^{-1}\[\fr{2}{\xi+R}-c\]
		- \coth^{-1}\[\fr{2}{\xi-R}-c\] 
	\end{align}
\end{subequations}

We verified by a straightforward calculation that our solution (2.28)
obeys the initial equations (2.7) at any value of the parameter $c$. 

It should be mentioned that we never assumed a separation of variables in 
solving the non-linear differential equations. However, when we choose 
$c=\pm1$ above, eqs.~(\ref{2.1.14a}) and (\ref{2.1.14b}) take the form
\begin{subequations}
	\begin{align}
		\xi&=\pm\[1-e^{\mp\eta}\cosh\chi\] \\
		R &= e^{\mp\eta}\sinh\chi 
	\end{align}
\end{subequations}
It is the solution that one easily gets by assuming a separation of variables, 
and it is precisely the one given in ref.~\cite{ll} --- see the footnote after 
eq.~(113.5) overthere. 

When choosing $c=0$, one gets a non-separable solution
\begin{subequations}
	\begin{align}
		\xi&=\fr{2\sinh\eta}{\cosh\eta +\cosh\chi}\\
		R &= \fr{2\sinh\chi}{\cosh\eta +\cosh\chi} 
	\end{align}
\end{subequations}
that can be found in refs.~\cite{two,three,four}. 

Thus our new solution (2.28) can also be considered as the interpolating 
solution between the previously known special solutions of refs.~\cite{ll} and
 \cite{two,three,four}. 

\section{Closed FRW case, $k=+1$}

A derivation of the coordinate transformation of the closed FRW metric
to its manifestly conformally flat form is almost the same as that in the 
previous section, with {\it sinh} and {\it cosh} being replaced by {\it sin}
 and {\it cos}, respectively. So, we skip many details here and present only
 our notation (in fact, very similar to that of Sec.~2) and our results.

First, we introduce a coordinate $\chi$ by
\begin{align}
	r = \sin \chi 
\end{align}
and some new coordinates  $\xi(\eta,\chi)$ and $R(\eta,\chi)$ in which the 
closed FRW
metric (1.2) would be manifestly conformally flat, 
\begin{align}
	-\mrd s^2 = {a}(\xi,R)A^2(\xi,R)
	\[ -\mrd\xi^2 + \mrd R^2 + R^2\mrd\Omega^2\] \lb{3.3}
\end{align}
It gives rise to the following set of the non-linear partial differential 
equations: 
\begin{subequations}
	\begin{align}
		-1 = A^2[-(\xi,_\eta)^2 + (R,_\eta)^2] \lb{3.7a}\\
		1 = A^2[-(\xi,_\chi)^2 + (R,_\chi)^2] \lb{3.7b} \\
		0 = -2 \xi,_\eta \xi,_\chi + 2 R,_\eta R,_\chi \lb{3.7c}\\
		\sin^2\chi = A^2(\xi,R)R^2 \lb{3.7d}.
	\end{align}
\end{subequations}

When proceeding as in the previous section, with the integrability condition
(\ref{2.13}) playing the central role, we arrive at a linear wave equation 
again,
\begin{align}
	\(
		\fr{\pr^2}{\pr \eta^2}
		- \fr{\pr^2}{\pr\chi^2}
	\)\xi
	=0 \lb{3.14}
\end{align}
whose general solution is
\begin{align}
	\xi=
	\xi_+(\eta+\chi) + \xi_-(\eta - \chi) \lb{3.1.1}
\end{align}
in terms of arbitrary functions $\xi_+$ and $\xi_-$ of $\eta+\chi$ and 
$\eta-\chi$, respectively. Similarly, one finds that
\begin{align}
	R=
	\xi_+(\eta+\chi) - \xi_-(\eta - \chi) \lb{3.1.4}
\end{align}

A substitution of eqs.~(\ref{3.1.1}) and (\ref{3.1.4}) into eq.~(\ref{3.7a}) 
yields
\begin{align}
	\fr{1}{\sin^2\chi}=\fr{4\xi'_+\xi'_-}{(\xi_+ -\xi_-)^2}
	\lb{3.1.5}
\end{align}
where the primes indicate the derivatives with respect to $\eta + \chi$ or 
$\eta -\chi$,
respectively. We again introduce the new coordinates $x$ and $y$ as follows:
\begin{subequations}
	\begin{align}
		x :=\eta+\chi&,~~y :=\eta-\chi  \lb{3.1.6} \\
		\chi=\fr{x-y}{2}&,~~\eta=\fr{x+y}{2}
	\end{align}
\end{subequations}and rewrite eq.~(\ref{3.1.5}) to the form
\begin{align}
	\fr{1}{\sin^2\fr{x-y}{2}}
	=\fr{4\xi'_+(x) \xi'_-(y)}{(\xi_+(x) -\xi_-(y))^2} \lb{3.1.7}
\end{align}
A solution to this equation is given by 
\begin{subequations}
	\begin{align}
		\xi_+(x) &=
		\fr{1}{\cot\fr{x}{2}+c_+} \lb{3.1.13a} \\ 
		\xi_-(x) &=
		\fr{1}{\cot\fr{y}{2}+c_-} \lb{3.1.13b}
	\end{align}
\end{subequations}with the same integration constants  $c_+=c_-:=c$. 
Here our main new results are ({\it cf.} eqs.~(\ref{2.1.14a}) and 
(\ref{2.1.14b}))
\begin{subequations}
	\begin{align}
		\xi
		&= \fr{1}{\cot\fr{\eta+\chi}{2}+c} + 
\fr{1}{\cot\fr{\eta-\chi}{2}+c} 
		\lb{3.1.14a}\\
		R
		&= \fr{1}{\cot\fr{\eta+\chi}{2}+c} - 
\fr{1}{\cot\fr{\eta-\chi}{2}+c} 
		\lb{3.1.14b}\\
		A^2(\eta,\chi)&= \fr{
		(\cot\fr{\eta+\chi}{2}+c)^2(\cot\fr{\eta-\chi}{2}+c)^2
		}{
		(\cot\fr{\eta-\chi}{2}-\cot\fr{\eta+\chi}{2})^2
		}\sin^2\chi 
		\lb{3.1.14c}
	\end{align}
\end{subequations}

The inverse transformation reads
\begin{subequations}
	\begin{align}
		\eta
		&=\cot^{-1}\[\fr{2}{\xi+R}-c\]
		+ \cot^{-1}\[\fr{2}{\xi-R}-c\] \\
		\chi
		&=\cot^{-1}\[\fr{2}{\xi+R}-c\]
		- \cot^{-1}\[\fr{2}{\xi-R}-c\]. 
	\end{align}
\end{subequations}

The transformations (\ref{3.1.14a}) and (\ref{3.1.14b}) appear to be 
{\it non-separable} for any {\it real} value 
of the parameter $c$, unlike the situation with $k=-1$ in the previous 
section. When 
(formally) taking $c=\pm i$, eqs.~(\ref{3.1.14a}) and (\ref{3.1.14b}) take the 
separable though complex form
\begin{equation} 
\xi=\pm\fr{1}{i}\[1-e^{\mp i \eta}\cos\chi\]~,\qquad
		R = e^{\mp i\eta}\sin\chi~, 
\end{equation}
i.e. there are no separable real solutions in the closed FRW case. 

When choosing $c=0$ in eqs.~(\ref{3.1.14a}) and (\ref{3.1.14b}), we get 
\begin{subequations}
	\begin{align}
		\xi&=\fr{2\sin\eta}{\cos\eta +\cos\chi}\\
		R &= \fr{2\sin\chi}{\cos\eta +\cos\chi} 
	\end{align}
\end{subequations}
thus reproducing the special transformation founded earlier in refs.~\cite{two,three,four}. 

\section{Conclusion}

The simplicity of our results is due to the fact that we were looking for the
relevant coordinate transformations in the two-dimensional space (or plane).
In physical terms, the null curves of that plane are to be invariant under such
transformations, while eq.~(\ref{2.1.7}) or (\ref{3.1.7}) is nothing but the 
invariance condition in the null coordinates (\ref{2.1.6}) or (\ref{3.1.6}),
respectively.

The real parameter $c$ entering eqs.~(2.28) and (3.11) just parametrizes the 
set of the coordinate transformations we found, so it does not have physical 
meaning. For instance, it does not appear in the standard (physically 
equivalent) form (1.1) or (1.2) of the FRW metric.

Once the FRW metric is transformed into the conformally flat form, there exist 
the 15-parametric group of four-dimensional conformal transformations (see 
e.g., ref.~(\cite{na})) that keeps the conformally flat form of the metric. 
Therefore, we can combine our one-parametric transformations in eqs.~(2.28) or
 (3.11) with the conformal transformations in four space-time dimensions, in 
order to get a much larger non-trivial 16-parametric family of the coordinate 
transformations bringing the standard $(k\neq 0$) FRW metrics to the 
manifestly conformally-flat form.
  
Though our results are rather technical, we believe that they may have 
interesting physical applications in cosmology and early Universe (see e.g.,
ref.~\cite{cosm}). The reason is that the FRW metrics are fully determined 
(up to a scale factor) by the symmetry, being independent upon equations of 
motion. 

All modern theories of quantum gravity, and especially string theory imply 
modifications of Einstein equations \cite{st,lg}. They are believed to be
crucial for any deeper understanding of inflation and Big Bang. Whatever 
those modifications are, they are going to include more fields and 
higher-curvature (or higher-derivative) terms in the effective gravitational
equations of motion, so that our considerations in this paper could be quite
useful for any such analysis at the level of the effective field equations.
Those equations include the full curvature, not just the Ricci tensor, so
that all the FRW curvature components are needed in their explicit form 
(see Appendix).

The 'non-flat' FRW  metrics in their manifestly conformally flat form may be 
particularly useful for applications of superstrings/M-theory to cosmology, 
because there are higher powers of curvature in the superstrings/M-theory 
effective field equations to all orders in the string slope parameter and the 
string coupling constant  \cite{st1,mine}, while the FRW curvatures take their 
simplest form just in such `conformally flat' coordinates. At the same time, 
it comes with the price: in the `conformally flat' coordinates the matter is 
not static anymore and is not even  homogeneous in general, though it still 
appears to be centrally-symmetric with respect to an arbitrary point in space 
(at the origin of the coordinate system).

\noindent {\it Note added:} after the submission of our paper to arXiv.org 
[hep-th/0702139] and for publication, we learned that the similar results 
also appeared in the later submission [gr-qc/0704.2788]. We thank M. Ibison 
for correspondence.

\section*{Acknowledgements}

One of authors (S.V.K.) would like to thank the Institute for Theoretical
Physics, Leibniz University of Hannover, Germany, for kind hospitality extended
to him during part of this investigation. This work is partially supported by 
the Japanese Society for Promotion of Science (JSPS) under the Grant-in-Aid 
programme for scientific research, and the bilateral german-japanese exchange
programme under the auspices of JSPS and DFG (Deutsche Forschungsgemeinschaft).

We are grateful to the referees of the Progress of Theoretical Physics journal 
for their careful reading of our manuscript and critical remarks.

\newpage

\section*{Appendix: FRW curvatures}

In this Appendix we calculate all the FRW curvatures, as well as the FRW Weyl 
tensor. To calculate the Riemann tensor $R_{\mu\nu ab}$, we choose to work with
 the spin connection $\omega_{\mu ab}$ and the {\it vierbein} $e^a_{\mu}$. We 
use lower case Greek letters for curved (spacetime) indices, and either lower 
case latin letters or lower case Greek letters with bars for flat (target 
space) indices.  The definitions are (see e.g., ref.~\cite{peter})
\begin{subequations}
	\begin{align}
		\omega_{\mu ab}=\frac{1}{2}e^c_{~\mu}\left( 
			\Omega_{cab}+\Omega_{bac}+
			\Omega_{bca}
		\right) \\
		\Omega_{abc}= e^{~\mu}_ae^{~\nu}_b\left(
			\partial_{\mu}e_{c \nu}-
			\partial_{\nu}e_{c\mu}
		\right)\\
		\Omega_{bac}=- \Omega_{abc}~,\quad \omega_{\mu ba}=-\omega_{\mu ab} \\
		R_{\mu\nu ab}= S_{\mu\nu ab}+K_{\mu\nu ab} \lb{5.4}\\
		S_{\mu\nu ab}= \partial_{\mu}\omega_{\nu ab}-\partial_{\nu}\omega_{\mu ab}
		\lb{5.5}\\
		K_{\mu\nu ab}= \omega_{\mu a}{}^c\omega_{\nu cb}-\omega_{\nu a}{}^c
		\omega_{\mu cb} \lb{5.6}
	\end{align}
\end{subequations}
To check their equivalence to the standard definition of Riemann curvature in 
terms of Christoffel symbols,
\begin{align}
	R^\mu_{~\nu\rho\gs}
	=\pr_\rho\gC^{\mu}_{\nu\gs}-\pr_\gs\gC^{\mu}_{\nu\rho}
	+\gC^\mu_{\rho\ga}\gC^{\ga}_{\nu\gs}
	-\gC^\mu_{\gs\ga}\gC^{\ga}_{\nu\rho} \lb{6.1}
\end{align}
where the Christoffel symbols are given by
\begin{align}
	\gC^\gl_{\mu\nu}=\fr{1}{2}g^{\gl\ga}\(
		g_{\ga\mu,\nu}+g_{\ga\nu,\mu}-g_{\mu\nu,\ga}\) 
\end{align}
it is most convenient to use the fully covariant constancy of the vierbein:
\begin{align}
	\pr_\mu e^a_{~\nu}
	+\gw_{\mu~b}^{~a}e^b_{~\nu}
	-\gC^\gl_{\mu\nu}e^a_{~\gl}
	=0 \lb{6.2}
\end{align}
and the fact that the definition of Riemann curvature in terms of the spin 
connection,
\begin{align}
	R_{\mu\nu~b}^{~~a}
	=\pr_\mu\gw_{\nu~b}^{~a}-\pr_\nu\gw_{\mu~b}^{~a}
	+\gw_{\mu~c}^{~a}\gw_{\nu~b}^{~c}
	-\gw_{\nu~c}^{~a}\gw_{\mu~b}^{~c} \lb{6.3}
\end{align}
is equivalent to a commutator of the covariant derivatives,
\begin{align}
	[D_\mu,D_\nu] V^a
	=R_{\mu\nu~b}^{~~a}V^b\lb{6.4}
\end{align}
where $D_{\mu}$ stands for the covariant derivative acting on local Lorentz 
(flat) indices only (in the vierbein formalism),
\begin{align}
	D_\mu e^a_{~\nu}
	=\pr_\mu e^a_{~\nu}+ \gw_{\mu~b}^{~a}e^b_{~\nu}
\end{align}

Then we have
\begin{align}
		D_\mu D_\nu V^a
		&=D_\mu D_\nu(e^a_{~\ga}V^\ga) \nonumber\\
	&=D_\mu[
		(
			D_\nu e^ a_{~\ga}
		)V^\ga
		+e ^a _{~\ga} D_\nu V^\ga
	] \nonumber\\
	&=D_\mu[
		\gC^\gb_{\nu\ga} e^a_{~\beta} V^\ga
		+e^a_{~\ga} \pr_\nu V^\ga
	]\nonumber\\
	&=\pr_\mu\gC^\beta_{\nu\ga} e^a_{~\beta}V^\ga
	+\gC^\beta_{\nu\ga}\gC^\gc_{\mu\beta}e^a_{~\gc}V^\ga
	+\gC^\beta_{\nu\ga}e^a_{~\gb}\pr_\mu  V^\ga \nonumber\\
	&~~+\gC^\gb_{~\mu\ga}e^a_{~\gb}\pr_\nu V^\ga
	+e^a_{~\ga}\pr_\mu\pr_\nu V^\ga
\end{align}
which gives rise to the commutator (\ref{6.4}) in the form
\begin{align}
	R_{\mu\nu~a}^{~~b}V^b
	&=	[D_\mu,D_\nu ]V^a \nonumber\\
	&=e^a_{~\gc} V^\ga[
		\pr_\mu\gC^\gc_{\nu\ga}
		+ \gC^\gb_{\nu\ga}\gC^\gc_{\mu\gb}
		-\pr_\nu\gC^\gc_{\mu\ga}
		-\gC^\gb_{\mu\ga}\gC^\gc_{\nu\gb}
	] \nonumber\\
	&=e^a_{~\gc} V^\ga R^\gc_{~\ga\mu\nu}	
\end{align}
Now the equivalence follows
\begin{align}
	R_{\mu\nu ab}
	=e_a^{~\ga}e_b^{~\gb}R_{\ga\gb\mu\nu}
	=e_a^{~\ga}e_b^{~\gb}R_{\mu\nu\ga\gb}
\end{align}
This equivalence is of course well-known, see e.g., ref.~(\cite{na}), so 
we consider our proof here as merely a consistency check of our notation.

When a metric is of the `diagonal' type
\begin{align}
ds^2=(A_0)^2(dz^0)^2- \sum_{\mu=1}^3 (A_{\mu})^2(dz^{\mu})^2
\end{align}
it is convenient to use a diagonal vierbein
\begin{align}
e^{\bar{\mu}}_{~\mu}=A_{\mu}
\end{align}

The FRW metric (1.2) is diagonal, with the components
\begin{subequations}
	\begin{align}
		g_\eta &:= g_{\eta\eta} = +a^2\\
		g_r&:=g_{rr} = -\frac{a^2}{1-kr^2}\\
		g_\theta &:= g_{\gq\gq} = -a^2r^2\\
		g_\phi &:= g_{\phi\phi} =-a^2r^2\sin^2\gq
	\end{align}
\end{subequations}
so that we have
\begin{align}
	A_\eta = a,\quad
	A_r = \frac{a}{\sqrt{1-kr^2}}~,\quad
	A_\gq = ar~,\quad
	A_\phi = ar\sin\gq
\end{align}
and
\begin{align}
	g_\mu = \eta_{\br{\mu}\br{\mu}}(A_\mu)^2
\end{align}
with the almost minus signature of our choice (as in ref.~\cite{ll}) 
\begin{align}
	\eta = \mr{diag}(+---)
\end{align}

The non-vanishing spin connection components of a diagonal metric are 
 (no summation over $\mu$ and $\nu$) 
\begin{align}
	\gw_{\mu\br{\mu}\br{\nu}} 
	= \eta_{\br{\mu}\br{\mu}}\frac{1}{A_\nu}\pr_\nu A_\mu \qquad
	 (\mu\neq\nu)
\end{align}
In the FRW case we have
\begin{align}
	\gw_{r\br{r}\br{\eta}}
	&=-\fr{1}{a}\dot{a}\fr{1}{\sqrt{1-kr^2}} \nonumber\\
	\gw_{\gq\br{\gq}\br{\eta}}
	&=-\fr{1}{a}\dot{a}r\nonumber\\
	\gw_{\gq\br{\gq}\br{r}}
	&= -\sqrt{1-kr^2} \nonumber\\
	\gw_{\phi\br{\phi}\br{\eta}}
	&=-\fr{1}{a}\dot{a}r\sin\gq \nonumber\\
	\gw_{\phi\br{\phi}\br{r}}
	&=-\sqrt{1-kr^2}\sin\gq \nonumber\\
	\gw_{\phi\br{\phi}\br{\gq}}
	&=- \cos\gq \nonumber
\end{align}
where the dots denote the derivatives with respect to $\eta$, 
\begin{align}
	\dot{a}:=a,_\eta
\end{align}

The Riemann curvature (\ref{5.4}) is a sum of eqs.~(\ref{5.5}) and (\ref{5.6}).
 The  $S_{\mu\nu\br{\rho}\br{\gs}}$ does not vanish when at least one of its 
indices $\rho$ and $\gs$ is either $\mu$ or $\nu$. First, we take $\rho = \mu$ 
and $\gs\neq\nu$. Then we find
\begin{subequations}
	\begin{align}
		S_{\mu\nu~~\gs}^{~~~\mu}
		&=\[
			(\pr_\nu{\ln A_\gs})(\pr_\gs\ln A_\mu)
			-(\pr_\nu{\ln A_\mu})(\pr_\gs\ln A_\mu)
			-(\pr_\nu\pr_\gs\ln A_\mu)
		\]\\
		S_{\mu\nu~~\nu}^{~~~\mu}
		&=\[
			(\pr_\nu\ln A_\nu)(\pr_\nu\ln A_\mu)
			-(\pr_\nu\ln A_\mu)(\pr_\nu\ln A_\mu)
			-(\pr_\nu\pr_\nu\ln A_\mu)
		\] \nonumber\\
		&+\eta^{\br{\mu}\br{\mu}}\eta_{\br{\nu}\br{\nu}}
		\fr{A^2_\nu}{A^2_\mu}\[
			(\pr_\mu\ln A_\mu)(\pr_\mu\ln A_\nu)
			-(\pr_\mu\ln A_\nu)(\pr_\mu\ln A_\nu)
			-(\pr_\mu\pr_\mu\ln A_\nu)
		\]
	\end{align}
\end{subequations}
Next we calculate $\pr_\nu(\ln A_\mu)$: 
\begin{align}
	(\ln A_\eta),_\eta 
	=(\ln a),_\eta \nonumber
\end{align}
\begin{align}
	(\ln A_r),_\eta
	&= (\ln a),_\eta \nonumber\\
	(\ln A_r),_r
	&=\fr{kr}{1-kr^2} \nonumber
\end{align}
then
\begin{align}
	(\ln A_\gq),_\eta
	&= (\ln a),_\eta \nonumber\\
	(\ln A_\gq), _r
	&=\fr{1}{r} \nonumber
\end{align}
and
\begin{align}
	(\ln A_\phi),_\eta
	&=(\ln a),_\eta \nonumber\\
	(\ln A_\phi),_r 
	&=\fr{1}{r} \nonumber\\
	(\ln A_\phi),_\gq
	&=\frac{1}{\tan\gq} \nonumber
\end{align}
All the other components of $\pr_\nu(\ln A_\mu)$ vanish.

Similarly, the components $S_{\mu\nu~~\gs}^{~~~\mu}$ (no sums!) are given by
\begin{align}
	S_{\gq r~~\eta}^{~~~\gq}
	=-\fr{1}{r}(\ln a),_\eta \nonumber
\end{align}
and
\begin{align}
	S_{\phi r~~\eta}^{~~~\phi} 
	&=-\fr{1}{r}(\ln a),_\eta \nonumber\\
	S_{\phi \gq~~\eta}^{~~~\phi}
	&=-\frac{1}{\tan\gq}(\ln a),_\eta \nonumber\\
	S_{\phi \gq~~r}^{~~~\phi}
	&=-\fr{1}{\tan\gq}\fr{1}{r} \nonumber
\end{align}
while otherwise zero.

To derive the remaining components $S_{\mu\nu~~\nu}^{~~~\mu}$ (no sums!), 
we define
\begin{align}
	B_{\mu\nu} 
	=(\ln A_\nu),_\nu(\ln A_\mu),_\nu
	-\[(\ln A_\mu),_\nu\]^2 -(\ln A_\mu),_{\nu},_{\nu}
\end{align}
The $S_{\mu\nu~~\nu}^{~~~\mu}$ can now be rewritten to
\begin{align}
	S_{\mu\nu~~\nu}^{~~~\mu}
	= B_{\mu\nu} 
	+ \eta^{\br{\mu}\br{\mu}}\eta_{\br{\nu}\br{\nu}}
	\fr{A_\nu^2}{A_\mu^2}B_{\nu\mu}
\end{align}
The non-vanishing components of $B_{\mu\nu}$ are
\begin{align}
	B_{r\eta}
	&=-(\ln a),_\eta,_\eta \nonumber\\
	B_{\gq\eta}
	&=-(\ln a),_\eta,_\eta \nonumber\\
	B_{\gq r}
	&=\fr{k}{1-kr^2} \nonumber\\
	B_{\phi\eta} 
	&=-(\ln a),_\eta,_\eta \nonumber\\
	B_{\phi r}
	&=\fr{k}{1-kr^2} \nonumber\\
	B_{\phi\gq}
	&=1 \nonumber
\end{align}
It is now easy to get
\begin{align}
	S_{\eta r~~r}^{~~~\eta}
	&=\fr{1}{1-kr^2}(\ln a),_\eta ,_\eta \nonumber\\
	S_{\eta \gq~~\gq}^{~~~\eta}
	&=r^2(\ln a),_\eta ,_\eta \nonumber\\
	S_{\eta \phi~~\phi}^{~~~\eta}
	&=r^2 \sin^2\gq(\ln a),_\eta ,_\eta \nonumber\\
	S_{r \gq ~~\gq}^{~~~r}
	&=r^2k \nonumber\\
	S_{r \phi~~\phi}^{~~~r}
	&=r^2\sin^2 k \nonumber\\
	S_{\gq \phi~~\phi}^{~~~\gq}
	&=\sin^2\gq \nonumber
\end{align}

As regards a derivation of the $K_{\mu\nu}{}^{\rho}{}_{\sigma}$ components from
eq.~(\ref{5.6}), they can be only non-zero when at least one of the indices 
$\rho$ or $\gs$  is either $\mu$ or $\nu$.

First, we set $\rho = \mu$ and $\gs\neq\nu$. Then we have
\begin{align}
	K_{\mu\nu~~\gs}^{~~~\mu}
	=(\pr_\nu\ln A_\mu)(\pr_\gs\ln A_\nu)
\end{align}
whose non-vanishing components are
\begin{align}
	K_{\gq r~\eta}^{~~\gq}
	=\fr{1}{r}(\ln a),_\eta \nonumber
\end{align}
and
\begin{align}
	K_{\phi r~\eta}^{~~\phi}
	&=\fr{1}{r}(\ln a),_\eta \nonumber\\
	K_{\phi\gq~\eta}^{~~\phi}
	&=\fr{1}{\tan\gq}(\ln a),_\eta \nonumber\\
	K_{\phi\gq~r}^{~~\phi}
	&=\fr{1}{\tan\gq}\fr{1}{r} \nonumber
\end{align}

Next, we set $\rho=\mu$ and $\gs=\nu$ in order to calculate 
$K_{\mu\nu}{}^{\mu}{}_{\nu}$ (no sums!)
\begin{align}
	K_{\mu\nu~~\nu}^{~~~\mu}
	=-\sum_{\gk\neq\mu,\nu}\eta^{\br{\gk}\br{\gk}}
	\eta_{\br{\nu}\br{\nu}}\fr{A_\nu^2}{A_\gk^2}
	(\pr_\gk\ln A_\mu)(\pr_\gk\ln A_\nu)
\end{align}
The non-vanishing components are given by
\begin{align}
	K_{r\gq~\gq}^{~~r}
	&=r^2 \[(\ln a),_\eta\]^2 \nonumber\\
	K_{r\phi~\phi}^{~~r}
	&=r^2\sin^2\gq \[(\ln a),_\eta\]^2 \nonumber\\
	K_{\gq\phi~\phi}^{~~\gq}
	&=r^2\sin^2\gq \[(\ln a),_\eta\]^2 
	-\sin^2\gq(1-kr^2) \nonumber
\end{align}

We are now in a position to get all the non-vanishing components of the Riemann
curvature from eq.~(\ref{5.4}). We find
\begin{subequations}
	\begin{align}
		R_{\eta r~r}^{~~\eta}
		&=\fr{1}{1-kr^2}(\ln a),_\eta,_\eta \\
		R_{\eta\gq~\gq}^{~~\eta}
		&=r^2(\ln a),_\eta,_\eta \\
		R_{\eta\phi~\phi}^{~~\eta}
		&=r^2\sin^2\gq(\ln a),_\eta,_\eta \\
		R_{r \gq ~\gq}^{~~r}
		&=r^2\[k + ((\ln a ),_\eta)^2\] \\
		R_{r\phi ~\phi}^{~~r}
		&=r^2\sin^2\gq\[k + ((\ln a ),_\eta)^2\] \\
		R_{\gq\phi ~\phi}^{~~\gq}
		&=r^2\sin^2\gq\[k + ((\ln a ),_\eta)^2\] 
	\end{align}
\end{subequations}

Accordingly, all the non-vanishing componentns of the Ricci tensor are given by
\begin{subequations}
	\begin{align}
		R_{\eta\eta}
		&=-3(\ln a ),_\eta,_\eta \\
		R_{rr}
		&=\fr{1}{1-kr^2}
		\[	(\ln a),_\eta,_\eta + 2k + 2((\ln a),_\eta)^2\] \\
		R_{\gq\gq}
		&=r^2
		\[(\ln a),_\eta,_\eta + 2k + 2((\ln a),_\eta)^2\] \\
		R_{\phi\phi}
		&=r^2\sin^2\gq
		\[	(\ln a),_\eta,_\eta + 2k + 2((\ln a),_\eta)^2\]
	\end{align}
\end{subequations}
Finally, the scalar curvature is
\begin{align}
	R=-\fr{6}{a^2}\[(\ln a),_\eta,_\eta + k + ((\ln a),_\eta)^2\]
\end{align}

The Weyl curvature tensor is the traceless part of the Riemann curvature 
tensor, and it is defined by
\begin{align}
	C_{\mu\nu~\gs}^{~~\rho}
	&=R_{\mu\nu~\gs}^{~~\rho} \nonumber\\
	&~~-\fr{1}{2}\(
		\delta^\rho_{~\mu}R_{\nu\gs}
		-\delta^\rho_{~\nu}R_{\mu\gs}
		-g_{\gs\mu}R_\nu^{~\rho}
		+g_{\gs\nu}R_\mu^{~\rho}
	\) \nonumber\\
	&~~+\fr{1}{6}R\(
		\delta^\rho_{~\mu}g_{\nu\gs}
		-\delta^\rho_{~\nu}g_{\mu\gs}
	\).
\end{align}
When using eqs.~(5.24), (5.25) and (5.26), it is easy to verify that all the 
Weyl tensor components do vanish in the case of the FRW curvature,
\begin{align}
	C_{\mu\nu~\gs}^{~~\rho}=0
\end{align}

To conclude the Appendix, a few comments are in order.

Of course, there are other ways to describe the vierbein formalism much shorter 
(e.g., when using the differential forms). Also, when using some deeper known 
results about physics and geometry, it may be possible to derive the Ricci tensor 
faster (e.g., when using its perfect fluid type, the rotational invariance, 
together with the Friedman and Raychaudhuri equations from the textbooks). However,
 all that would not help in a calculation of the FRW curvature components, which 
was one of our main objectives in this paper. 
 
The fact of the vanishing Weyl curvature also follows from the standard (Petrov) 
classification \cite{petrov} of curved spacetimes: the local isotropy is allowed by 
the Petrov types D and N only, however the symmetries of those types do not include 
spacial rotations. So, our explicit check of the vanishing Weyl tensor may be 
considered as a non-trivial check of our calculations of the FRW curvatures.

\end{document}